\documentclass{aa}  

\usepackage{graphicx}
\usepackage{txfonts}
\usepackage{float}
\usepackage[colorlinks=true, citecolor=blue]{hyperref}
\usepackage[dvipsnames]{xcolor}
\usepackage{soul}

\newcommand{\etass}{\eta^\mathrm{sec}_\mathrm{so}}

\newcommand{\etat}{\eta_\mathrm{t}}
\newcommand{\etaap}{\eta_\mathrm{ap}}

\begin{document} 

   \title{Optimisation-based alignment of wideband integrated superconducting 
spectrometers for sub-mm astronomy}

   \author{A. Moerman\fnmsep\thanks{Corresponding author \email{A.Moerman-1@tudelft.nl}}
          \inst{1}
          \and
          K. Karatsu\inst{1,2}
          \and
          S. J. C. Yates\inst{3}
          \and
          R. Huiting\inst{2}
          \and
          F. Steenvoorde\inst{4}
          \and
          S. O. Dabironezare\inst{1}
          \and
          T. Takekoshi\inst{5}
          \and
          J. J. A. Baselmans\inst{1,2}
          \and
          B. R. Brandl\inst{6,7}
          \and
          A. Endo\inst{1}
          }
   
    \institute{Faculty of Electrical Engineering, Mathematics and Computer Science, Delft University of Technology, Mekelweg 4, 2628 CD, Delft, The Netherlands
         \and    
    SRON—Netherlands Institute for Space Research, Niels Bohrweg 4, 2333 CA, Leiden, The Netherlands     
        \and     
    SRON—Netherlands Institute for Space Research, Landleven 12, 9747 AD, Groningen, The Netherlands
        \and
    DEMO: Electronic and Mechanical Support Division, Delft University of Technology, Mekelweg 4, 2628CD, Delft, The Netherlands
        \and
    Kitami Institute of Technology, 165 Koen-cho, Kitami, 090-8507 Hokkaido, Japan
        \and
    Leiden Observatory, Leiden University, PO Box 9513, 2300 RA, Leiden, The Netherlands
        \and
    Faculty of Aerospace Engineering, Delft University of Technology, Kluyverweg 1, 2629 HS, Delft, The Netherlands\\
             }
   
    \date{Received; accepted}


    \abstract
    {Integrated superconducting spectrometers (ISSs) for wideband sub-mm astronomy utilise quasi-optical systems for coupling radiation
    from the telescope to the instrument. Misalignment in these systems is detrimental to the system performance. The common method of using an
    optical laser to align the quasi-optical components requires accurate alignment
    of the laser to the sub-mm beam coming from the instrument, which is not
    always guaranteed to a sufficient accuracy.}
    {To develop an alignment strategy for wideband ISSs directly utilising
    the sub-mm beam of the wideband ISS. The strategy should be applicable
    in both telescope and laboratory environments. Moreover, the strategy should
    deliver similar quality of the alignment across the spectral range of the wideband
    ISS.}
    {We measure misalignment in a quasi-optical system operating at
    sub-mm wavelengths using a novel phase and amplitude measurement
    scheme, capable of simultaneously measuring the complex beam patterns of a direct-detecting ISS across a harmonic range of frequencies. The direct detection nature of the MKID detectors in our device-under-test, DESHIMA 2.0,
    necessitates the use of this measurement scheme. Using
    geometrical optics, the measured misalignment, a mechanical hexapod, and an optimisation algorithm,
    we follow a numerical approach to optimise the positioning of corrective optics with respect to a given cost function. Laboratory measurements of the complex beam patterns
    are taken across a harmonic range between 205 and 391 GHz and simulated
    through a model of the ASTE telescope in order to assess the performance of
    the optimisation at the ASTE telescope.}
    {Laboratory measurements show that the optimised optical setup corrects for tilts and offsets of the sub-mm beam. Moreover, we find that the simulated telescope aperture efficiency is increased across the frequency
    range of the ISS after the optimisation.}
   {}


   \maketitle
%

\section{Introduction}
\label{sect:intro}  
Wideband sub-mm spectroscopy could serve as a powerful tool for studying a wide range of astrophysical phenomena~\citep{Stacey_2011}. For single-pixel spectroscopy, one such target is the redshifted [CII] emission line, which can be used to probe star formation over cosmic time~\citep{Lagache_2018}, study the universe at the epoch of reionization \citep{Gong_2011}, and study high-redshift dusty star-forming galaxies~\citep{Rybak_2022}. Multi-pixel spectrometers, also called integral field units (IFUs), can spectroscopically observe wide field-of-views. This allows for studies on larger spatial scales, such as line intensity mapping of the [CII] line~\citep{Yue_2015,Yue_2019,Karoumpis_2022} which can be used to study the growth of large-scale structure (LSS) in the early Universe. Also, the extragalactic rotational emission lines of CO can be used in cross-correlation power spectra studies with other LSS tracers such as the CIB~\citep{Maniyar2023} and the Ly-$\alpha$ forest signal~\citep{Qezlou2023}.

State-of-the-art integrated superconducting spectrometers (ISSs) for ground-based wideband sub-mm astronomy, such as Superspec~\citep{Karkare2020}, µ-spec~\citep{Mirzaei2020} and DESHIMA~\citep{Endo_2019B}, rely on superconducting detectors called microwave kinetic inductance detectors~\citep{Day2003,Baselmans_2012} (MKIDs) to detect incoming radiation. Unlike quasi-optical spectrometers such as CONCERTO~\citep{Ade_2020}, an ISS integrates both the detectors and the dispersive element on a single chip. This includes everything from the antenna capturing the incoming radiation, the filterbank or dispersive element that separates spectral channels, down to the MKID detectors. An obvious advantage of the ISS is the scalability from a single-pixel instrument to a multi-pixel IFU: because the entire device is already fabricated on a single chip, it is possible to fabricate multiple of these devices on a single wafer.

Coupling the broadband single-mode signal from the telescope to the single-pixel ISS requires good alignment of all intermediate optical components. A simple and effective design that is often adopted for heterodyne receivers (see for example the ALMA band 5, 8, and 9 receivers~\citep{Belitsky2018,Satou2008,Baryshev2015}) is to rigidly mount the waveguide feed horn of the SIS mixer~\citep{Tucker1985} (and optionally a simple 4 K fore-optics for polarization separation) near the Cassegrain focus of the telescope, relying only on the movement of the secondary mirror to adjust the focus. The requirements on the cartridges and beam pointings are strict, ensuring tight mechanical tolerances. However, placing the detectors in the vicinity of the telescope focus is not always possible, in which case intermediate mirrors are introduced. For example, switching mirrors might be introduced for multiple instruments to share the same focus. Especially in the case of instruments with direct-detection detectors that operate at sub-Kelvin temperatures, the cold detector mount tends to be mechanically more complex, and there can be a set of cold re-imaging optics to reject stray light and out-of-band thermal influx~\citep{Lamb2003,Holland2013,Endo_2019A}. Each of these optical components have a finite alignment tolerance, and make it hard to achieve good total optical coupling by adjusting only the secondary mirror of the telescope. Additionally, errors in the mounting of the cryostat housing the ISS introduce extra alignment issues, furthering the problem.

One approach to correct for the misalignment in the quasi-optical chain is to make one or more mirrors adjustable in their position. However, finding the optimum position of these components in a systematic and reproducible manner is not a trivial problem in practice. A common method is to use a visible-light laser~\citep{Catalano2022,Endo_2019B}, but it requires the propagation axis of the sub-mm beam to be aligned well with the laser ray, which is not always guaranteed. Moreover, the method does not offer a way to verify the quality of alignment in the telescope, nor does it give focal shifts.

Here we demonstrate a method to adjust the position of an optical component to correct for misalignment in a sub-mm quasi-optical system, which utilises a pair of reflectors mounted on a motorized hexapod, a novel phase-and-amplitude beam-pattern measurement technique, and an optimisation algorithm. We apply the alignment method to the DESHIMA 2.0~\citep{Taniguchi_2022} sub-mm ISS. DESHIMA 2.0 utilises a wideband leaky-lens antenna~\citep{Neto2010a,Neto2010b,Hahnle2020}, in combination with a superconducting filterbank~\citep{PascualLaguna2021,Thoen2022} coupled to an array of NbTiN-hybrid MKID~\citep{Janssen2013} detectors, to observe the electromagnetic spectrum between 220 and 440 GHz. The MKIDs are read out using a frequency-division multiplexed readout~\citep{vanRantwijk2016}, which allows for simultaneous readout of all spectral channels. The DESHIMA 1.0 instrument has seen first light~\citep{Endo_2019A} at the ASTE~\citep{Ezawa_2004} telescope, where a 4$^\mathrm{\circ}$ residual beam tilt after the laser alignment was established as a cause for the low aperture efficiency and distorted far-field beam patterns during the observations. This makes its successor, DESHIMA 2.0, a prime candidate for the development of the alignment strategy.

The paper is structured as follows: in Sect.~\ref{subsec:optics} we start by briefly introducing the optical chain of DESHIMA 2.0: the cryostat optics, the cabin optics and the ASTE Cassegrain reflector. In Sect.~\ref{subsec:HPA} we discuss the measurement technique we employ to obtain phase-amplitude beam patterns of the ISS across a harmonic range of signal frequencies simultaneously.
In Sect.~\ref{subsec:misalign}, we discuss how we can use this measurement technique to extract misalignment of the beam with respect to the designed optical axes of the system. In Sect.~\ref{subsec:optimisation}, we discuss the optimisation strategy and how this numerical approach can be applied to the alignment of a real-world optical system.  In Sect.~\ref{subsec:expsetup} we give a description of the experimental setup for the alignment procedure.
In Sect.~\ref{subsec:reduction} we present results from laboratory measurements to illustrate the performance of the optimised optical setup. In Sect.~\ref{subsec:efficiency}, the complex beam patterns obtained from the measurements are simulated through the ASTE telescope to assess the performance and efficiencies of the aligned ISS at the telescope. Lastly, in Sect.~\ref{subsec:ff}, we qualitatively investigate the far-field patterns of the instrument.

\section{Methods}
\label{sect:methods}
\subsection{Optical chain}
\label{subsec:optics}
\begin{figure*}[hbt!]
\centering
    \includegraphics[width=0.95\textwidth]{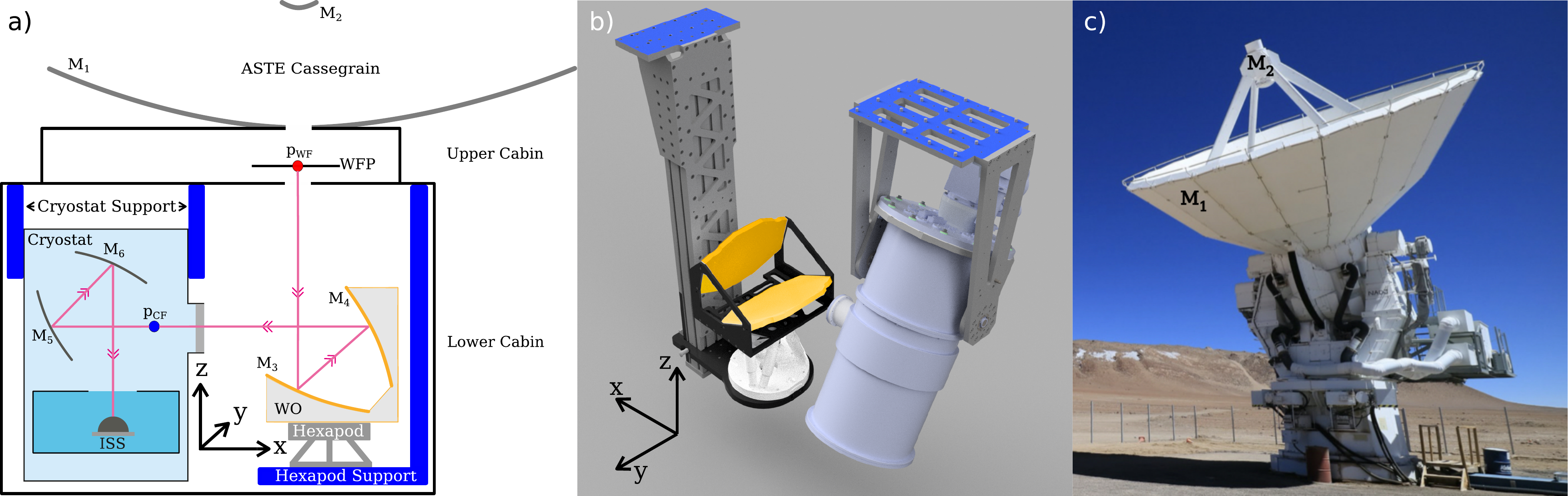}
    \caption{a) Sketch of the cryostat optics, warm optics and ASTE Cassegrain setup. In addition to the optical elements, the hexapod, cryostat, support structures and ASTE lower/upper cabin are also sketched. The warm focal plane is denoted `WFP' and the warm optics `WO' in this sketch. The drawing is not to scale. Note the Cartesian coordinate basis illustrated in between the cryostat and warm optics. This is the standard coordinate system for the rest of this work. 
    b) Render of the lower cabin setup at ASTE. the warm optics is illustrated in orange. The cryostat and hexapod support structures are attached to the lower cabin ceiling at the blue surfaces. For reference, we show the coordinate system from a) again.
    c) Photograph of the ASTE telescope in the Atacama desert. The primary reflector $M_1$ and secondary reflector $M_2$ are indicated. The lower cabin is located underneath the primary mirror in between the mounting fork.}
    \label{fig:setup_WO}
\end{figure*}

We begin by discussing the elements present in the optical path of DESHIMA 2.0 at ASTE. For a graphical overview of the optical setup, see, for example, Fig. 5.2 of~\citep{Dabironezare}. We recreate the setup in Fig.~\ref{fig:setup_WO} for completeness. We take the coordinate system given in Fig.~\ref{fig:setup_WO} as our standard: whenever we refer to an $x$, $y$ or $z$-coordinate in this work, we refer to this system.

The first optical element encountered by radiation coming from the sky is the Cassegrain setup at ASTE. The primary reflector $M_1$ is a parabolic reflector that forms a Cassegrain setup together with $M_2$, a hyperbolic secondary reflector. The Cassegrain setup directs the light through the upper cabin, where the Cassegrain focus is located, into the lower cabin. The lower cabin houses a dual reflector, which is based on a Dragonian~\citep{Dragone_1978} reflector design. The dual reflector acts as a coupler between the Cassegrain setup and the cryostat housing of DESHIMA 2.0 and can conveniently act as correcting optics when attached to the mechanical hexapod. The dual reflector and hexapod are attached to the lower cabin ceiling by a support structure. The first dual reflector mirror in the chain, $M_3$, is an off-axis ellipsoid. The second mirror, $M_4$, is an off-axis hyperboloid and couples light from $M_3$ into the cryostat. Because this dual reflector is situated outside the cryostat, it is referred to as the `warm optics'. 

One focus of the warm optics coincides with the Cassegrain focus. This focus is located above the warm optics (see Fig.~\ref{fig:setup_WO}a) and is called the `warm focus' ($\Vec{p}_\mathrm{WF}$). The plane parallel to the $xy$-plane, containing $\Vec{p}_\mathrm{WF}$, is called the `warm focal plane'. The second warm optics focus is located to the left of the warm optics. This focus, called the `cold focus' ($\Vec{p}_\mathrm{CF}$), is located inside the cryostat and coincides with the first focus of the cryostat optics. Our coordinate system is placed such that the origin lies in $\Vec{p}_\mathrm{CF}$. The plane parallel to the $yz$ plane and containing $\Vec{p}_\mathrm{CF}$ will be denoted the `cold focal plane'. The cryostat optics consists of a parabolic relay~\citep{Dabironezare} with two off-axis paraboloid reflectors, $M_5$ and $M_6$, in an aberration compensating configuration~\citep{Murphy1987}. The cryostat optics couple the radiation entering the cryostat to the ISS, which is located below the cryostat optics in the second focus.
The lower cabin ceiling acts as a mechanical reference for the placement of the entire optical setup inside the lower cabin. Therefore, any mounting errors of the cryostat and hexapod support structures amounts to misalignment of the system with respect to the entire telescope. 

The final element in the optical chain is the leaky-lens antenna mounted on the DESHIMA 2.0 chip. This structure converts the radiation impinging on the lens into guided radiation which is fed to the filterbank through a co-planar waveguide.

\subsection{Harmonic phase-amplitude measurements}
\label{subsec:HPA}
\begin{figure*}[hbt!]
\centering
    \includegraphics[scale=0.39]{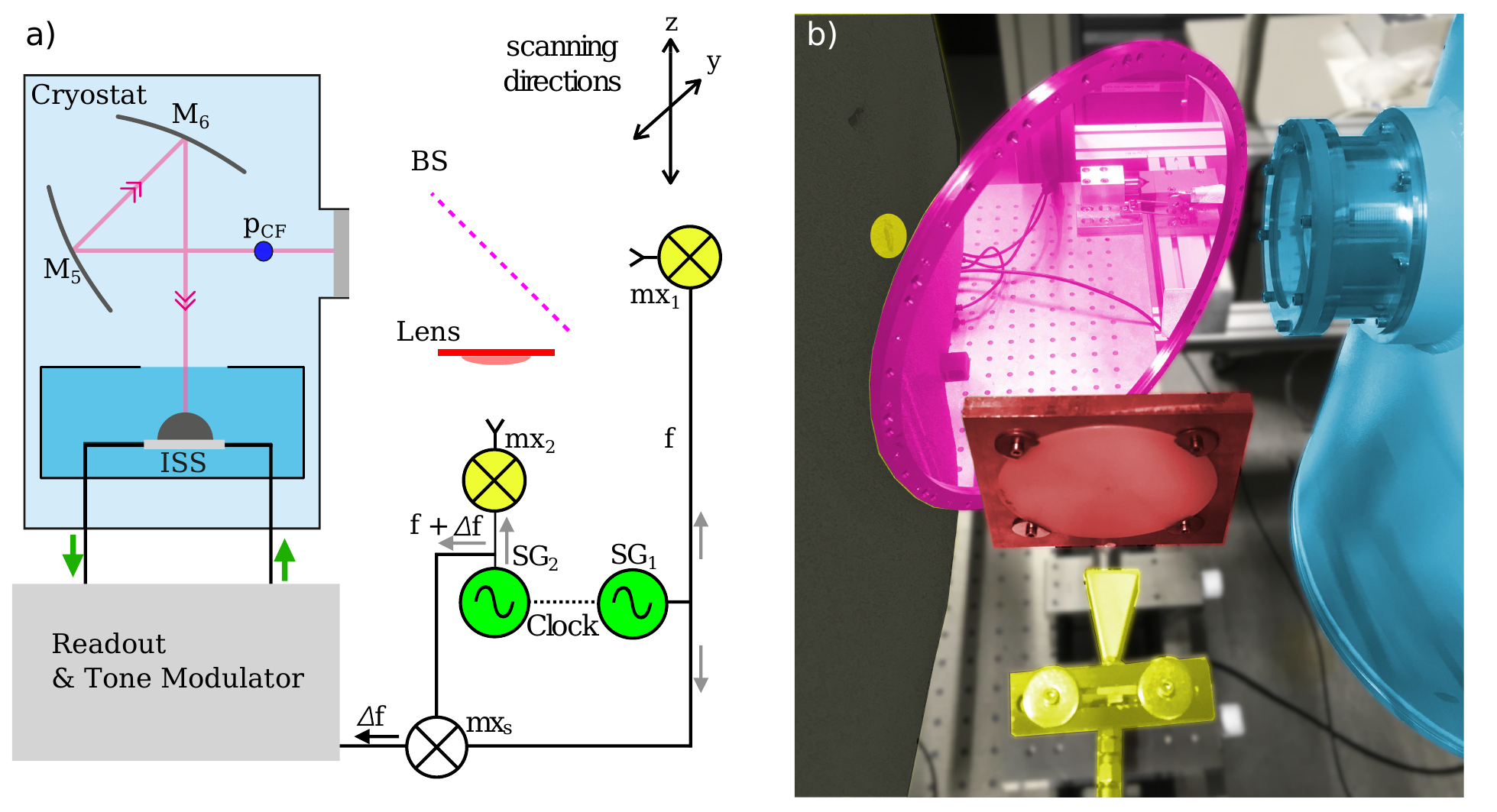}
    \caption{a) Sketch of the cryostat and HPA measurement setup. In green, the two signal generators SG$_1$ and SG$_2$. The harmonic mixers are depicted in yellow. The beamsplitter, colored pink and denoted BS, is situated such that its faces are illuminated by both mx$_1$ and mx$_2$. A lens, depicted in red, is placed in between mx$_2$ and the beamsplitter to increase the coupling. The modulated signal $U_\mathrm{m}$ then travels through the cryostat to the ISS. b) Photograph of the HPA measurement setup in front of the cryostat window. The components are color coded in the same fashion as in a). The mixer mx$_1$ is located inside the yellow disk on the left. The material in which mx$_1$ is embedded is a sheet of radiation absorbing material. This prevents standing waves from generating in between mx$_1$ and the cryostat window. In this photograph, the $x$-axis depicted in Fig.~\ref{fig:setup_WO}a is pointing out of the cryostat window, into the beamsplitter and scanning plane. The scanning plane is oriented along the $z$ and $y$-axes. The diagonal horn containing mx$_2$ is also colored yellow and visible in the bottom of the image, below the lens which is colored red.}
    \label{fig:setup_HPA}
\end{figure*}

In order to assess the performance of the optimisation strategy, we need a method to calculate how our measured beam patterns would propagate through the telescope in which the device will be mounted. Coherent propagation of electromagnetic fields, taking into account diffraction from the telescope reflectors, is only possible when both the phase and amplitude (PA) patterns of the instrument are known. In addition, misalignment of the sub-mm beam can be extracted from the PA patterns. We therefore employ a quasi-heterodyne measurement technique~\citep{Davis_2019,Yates_2020} which allows us to obtain PA beam patterns of direct detectors, such as MKIDs. We specifically employ an extension to the measurement technique, called the harmonic phase and amplitude (HPA) measurement. This novel measurement technique is capable of mapping PA beam patterns across a harmonic range of sub-mm frequencies simultaneously. See Fig.~\ref{fig:setup_HPA} for a graphical overview and the laboratory setup for the HPA measurement.

The HPA measurement technique relies on the temporal modulation of an incoming sub-mm radiation field $U_1$ by another sub-mm field $U_2$. The field $U_1$ is generated by feeding a synthesizer signal SG$_1$ at a frequency $f_0 = 9.77$ GHz to a harmonic mixer mx$_1$. We use an ultra-wideband superlattice mixer~\citep{Paveliev2012}, mounted in an open-ended wr4.3\footnote{A wr4.3 to wr2.2 transition and wr2.2 corrugated horn was used as a crosscheck at higher frequencies, so to work as a single moded source (as wr4.3 allows higher order modes above ~260GHz), but no difference was seen in the measured results (except higher S/N, as the efficiency of higher harmonics increased).} waveguide. This mixer generates overtones at integer multiples of $f_0$ and radiates the field through the waveguide into free space, resembling a point source in its far-field~\citep{Yaghjian1984}. The mixer mx$_1$ can move in a plane, the scanning plane, which allows us to map the instrument beam pattern as a function of mx$_1$ position. 

\begin{figure}[hbt!]
\centering
    \includegraphics[width=0.5\textwidth]{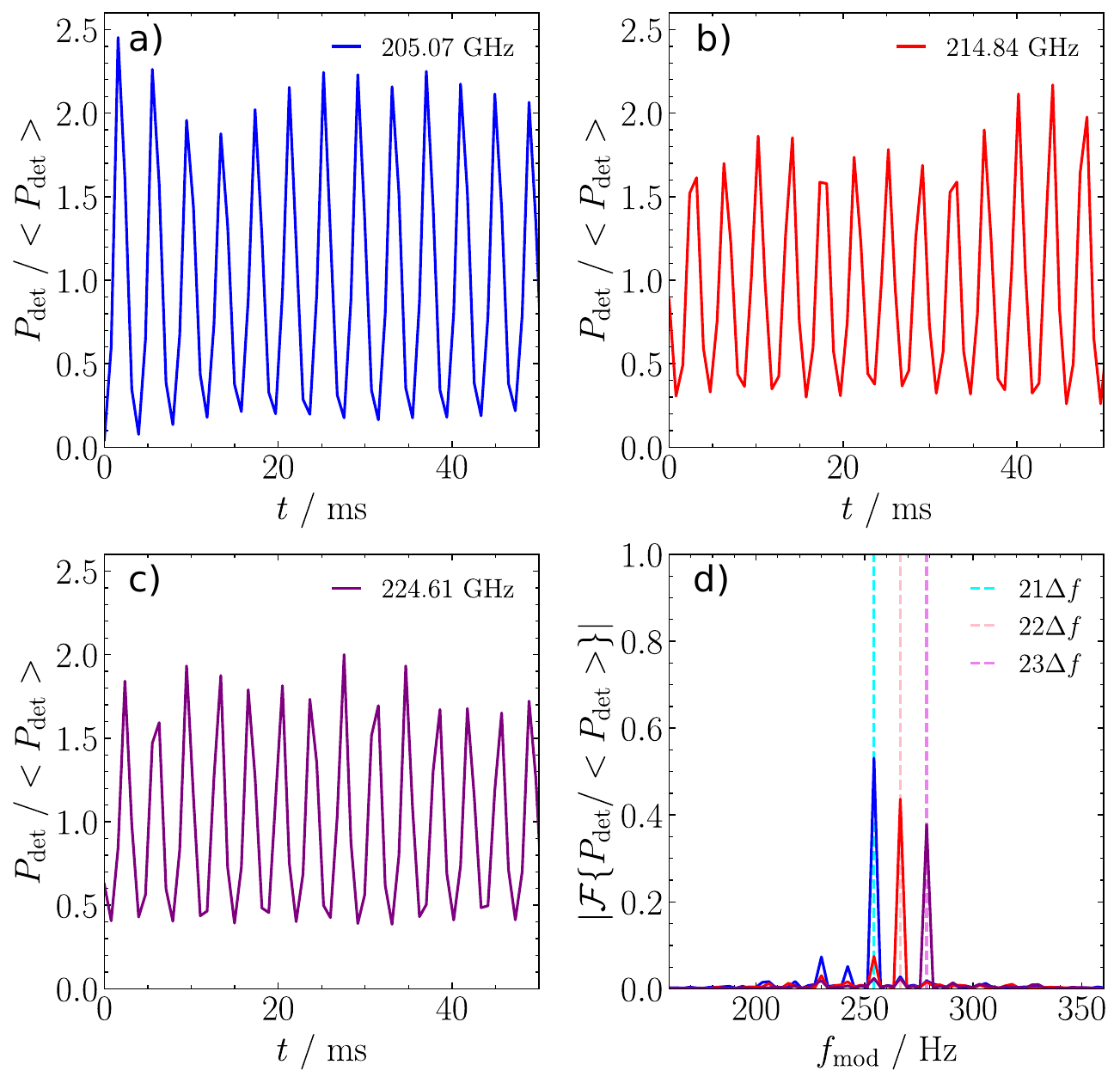}
    \caption{a) raw timestream data for $P_\mathrm{det}$ of MKID 171 at 205.07 GHz, taken at the center of the scanning plane. We have normalised $P_\mathrm{det}$ to the time-averaged power $<P_\mathrm{det}>$ entering the MKID. We plot the data for a time range of 50 ms. The modulation at $n \Delta f$ is visible as the rapid oscillation of the timestream data, which has a period of about 3.9 ms. b) raw timestream of MKID 96 at 214.84 GHz, taken at the center of the scanning plane. c) raw timestream of MKID 47 at 224.61 GHz, taken at the center of the scanning plane. d) Magnitude of the Fourier transforms of the signals in a), b), and c). The abscissa represents the frequencies present in the time-modulation of $P_\mathrm{det}$ and is denoted by $f_\mathrm{mod}$. The cyan dashed vertical line represents the $n\Delta f$ for $nf_0 = 205.07$ GHz. Given the value of $f_0$, this corresponds to $n=21$ and thus $n\Delta f = 254.31$ Hz, which is where we would expect the peak. The pink dashed line corresponds to $n=22$ for $nf_0=214.84$ GHz and the violet dashed line to $n=23$, $nf_0=224.61$ GHz.}
    \label{fig:time_fft}
\end{figure}

The field $U_2$ is generated by another harmonic mixer mx$_2$, identical to mx$_1$ but fed by another synthesizer SG$_2$, at a frequency $f_0 + \Delta{f}$. The mixer mx$_2$ is fixed spatially and radiates a beam with a low opening angle into free space through a diagonal horn antenna.
SG$_1$ and SG$_2$ generate their signals on a common clock, phase-locking $U_1$ and $U_2$ together. We choose $\Delta{f} = 12.11\:\mathrm{Hz}<f_r$, where $f_r = 1.2$ kHz is the sampling frequency of the MKIDs of the spectrometer. Then $U_1(t) = \sum_n A_n \sin(2\pi n f_0 t + \phi_{n,1})$, with $n \in \{21,...,41\} \subset \mathbb{N}$ in our setup. This corresponds to a frequency range between 205.07 and 390.62 GHz. The mixer mx$_2$ radiates $U_2(t) = \sum_n B_n \sin(2\pi n (f_0 + \Delta{f}) t + \phi_{n,2})$. These two fields radiate into free space and illuminate a mylar beamsplitter, where they are added together to generate $U_\mathrm{m} = U_1 + U_2$, the modulated field. In order to increase the coupling of $U_2$ to the MKIDs, we place a lens between mx$_2$ and the beamsplitter. Because $U_\mathrm{m}$ is amplitude-modulated at integer multiples of $\Delta{f}$ and $f_0$, the detected power $P_\mathrm{det} \propto |U_\mathrm{m}|^2$ in the instrument is modulated at multiple modulation frequencies, with each modulation frequency corresponding to a different harmonic contained in $U_1$:
\begin{equation}
    \label{eq:modulation}
    P_\mathrm{det}(t) \propto \sum_n A_n B_n \cos(2\pi n \Delta{f} t - \phi_n).
\end{equation}
Where $\phi_n = \phi_{n,1} + \phi_{n,2}$, the phase of the modulation frequency $n \Delta{f}$ of $P_\mathrm{det}$. We have dropped all modulation terms that have a frequency higher than $f_r$, as these terms do not contribute to the measurable time-dependence of $P_\mathrm{det}$ but contribute to the constant power entering the MKID. 
The amplitude $A_n B_n$ for each modulation frequency $n \Delta f$ can be extracted by taking the Fourier transform of $P_\mathrm{det}(t)$, for each point in the scanning plane, and finding the amplitudes corresponding to each $n \Delta f$. An example of this is illustrated in Fig.~\ref{fig:time_fft}, where we show the raw timestream data and Fourier transforms for three different MKIDs with filters at $nf_0$, each simultaneously recorded at a single point in the scanning plane.

We utilise a phase reference signal to extract $\phi_{n,1}$ from $\phi_n$. This signal is generated from SG$_1$ and SG$_2$ by feeding both signals into a subtractive mixer mx$_\mathrm{s}$. The resulting signal, after low-pass filtering, has frequency $\Delta{f}$ and some phase $\phi_\mathrm{ref}$.
We electronically modulate the entire instrument readout with the phase reference signal. Apart from tones coupled to MKIDs, the readout contains several `blindtones' that are not coupled to any MKIDs, in order to monitor temporal drift during observations. In post-processing, we can extract the phase reference signal, and hence $\phi_\mathrm{ref}$, by taking the Fourier transform of the blindtone readout timestreams and identifying the amplitude and phase of the Fourier transform at $\Delta f$, for each point in the scanning plane. After extraction, we can use the same blindtones to remove the readout modulation from the MKID-coupled tones, which modulate at a different frequency, to further suppress any leakage of the phase reference.
By referencing the phases $\phi_n$ of $P_\mathrm{det}$ to the phase reference signal phase $\phi_\mathrm{ref}$ at each source position in the scanning plane, we can spatially map the PA beam patterns across the harmonic range indexed by $n$.

\subsection{Measuring misalignment}
\label{subsec:misalign}
We call a beam, defined on some plane $P=P(x,y,z) \subset \mathbb{R}^3$ with beam amplitude center at some point $\Vec{p} \in P$, propagating in some direction $\Vec{\hat{\beta}}$, misaligned with respect to an axis $\Vec{\hat{\gamma}}$ if:

\begin{align}
&\Vec{p} \notin \mathrm{span}(\Vec{\hat{\gamma}}), \label{expr:pos}\\
&\Vec{\hat{\beta}} \cdot \Vec{\hat{\gamma}} < 1. \label{expr:tilt}
\end{align}

Expression~\ref{expr:pos} corresponds to beam offsets, where the $P-\Vec{\hat{\gamma}}$ intersection does not coincide with $\Vec{p}$. Expression~\ref{expr:tilt} corresponds to beam tilt, where the propagation direction of the beam does not coincide with the specified axis.

We measure beam misalignment at the cryostat by performing an HPA measurement in front of the cryostat window (see Fig.~\ref{fig:setup_HPA}b for the laboratory setup). The beam misalignment is with respect to the optical axis coming out of the cryostat window, which corresponds to the $x$-axis. The optical axis originates from the cold focus $\Vec{p}_\mathrm{CF}$. The beam propagation direction coming out of the cold focal plane is denoted $\Vec{\hat{\beta}}'_B$.

\begin{figure}[hbt!]
\centering
    \includegraphics[width=0.5\textwidth]{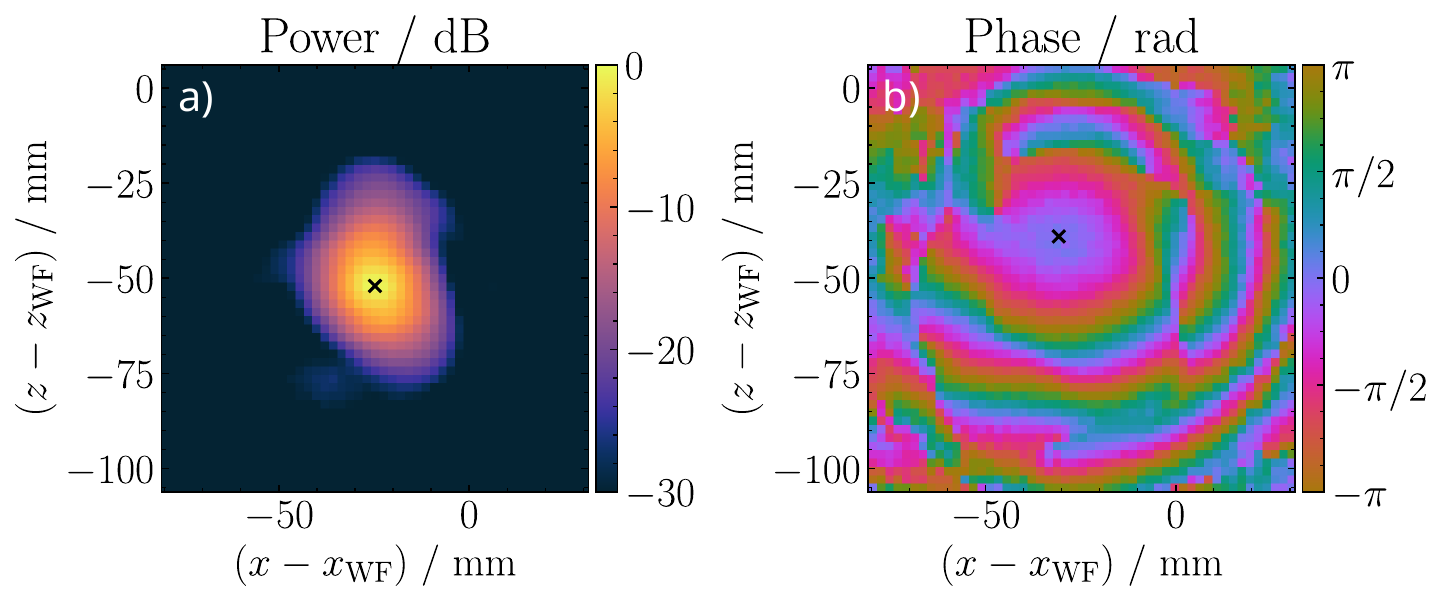}
    \caption{Complex beam patterns at 205 GHz, as measured with the HPA technique after the warm focal plane, with the hexapod in home configuration. a) Amplitude pattern. b) Phase pattern. The co-ordinates are with respect to the warm focus. Note that, since these are measured in the laboratory using the setup in Fig.~\ref{fig:HPA_WO}b, we use the scanning plane axes as shown in Fig.~\ref{fig:HPA_WO}a, with the warm focus in the $xz$-plane. We denote the amplitude and phase centers with black crosses. It can be seen that these are different, which indicates that the beam is tilted with respect to the scanning plane normal.}
    \label{fig:beamtilt}
\end{figure}

Before we start the measurement, we center mx$_1$ on the beam coming from the ISS, by finding the position in the scanning plane at which the MKID responses are the highest. Then, we start the HPA measurement and the beam patterns are obtained, one for each frequency $nf_0$. We fit a complex-valued, astigmatic Gaussian beam to each measured beam pattern. This fit can be used to obtain $\Vec{\hat{\beta}}'_B$~\citep{Tong2003,Davis_2016} by also including the rotation of the plane of the Gaussian as a free parameter in the fit. This is possible because beam tilt introduces a mismatch between the amplitude and phase centers of the beam pattern in the scanning plane~\citep{MingTangChen}. See Fig.~\ref{fig:beamtilt} for an illustration of this. Additionally, the Gaussian fit produces the location of the fitted Gaussian beam focal spot with respect to the scanning plane center, which we will assume to be equal to $\Vec{p}_\mathrm{CF}$, the cryostat focus of the actual beam.

The hexapod and warm optics setup could also contribute misalignment due to, for example, mounting errors of the support structure. In this case, the optical axis is the $z$-axis and the point we take as reference is the warm focus $\Vec{p}_\mathrm{WF}$. 
The beam center in the warm focal plane is denoted $\Vec{p}_B$. The beam propagation direction coming out of the warm focal plane is denoted by $\Vec{\hat{\beta}}_B$.

\begin{figure*}[h]
\centering
    \includegraphics[scale=0.39]{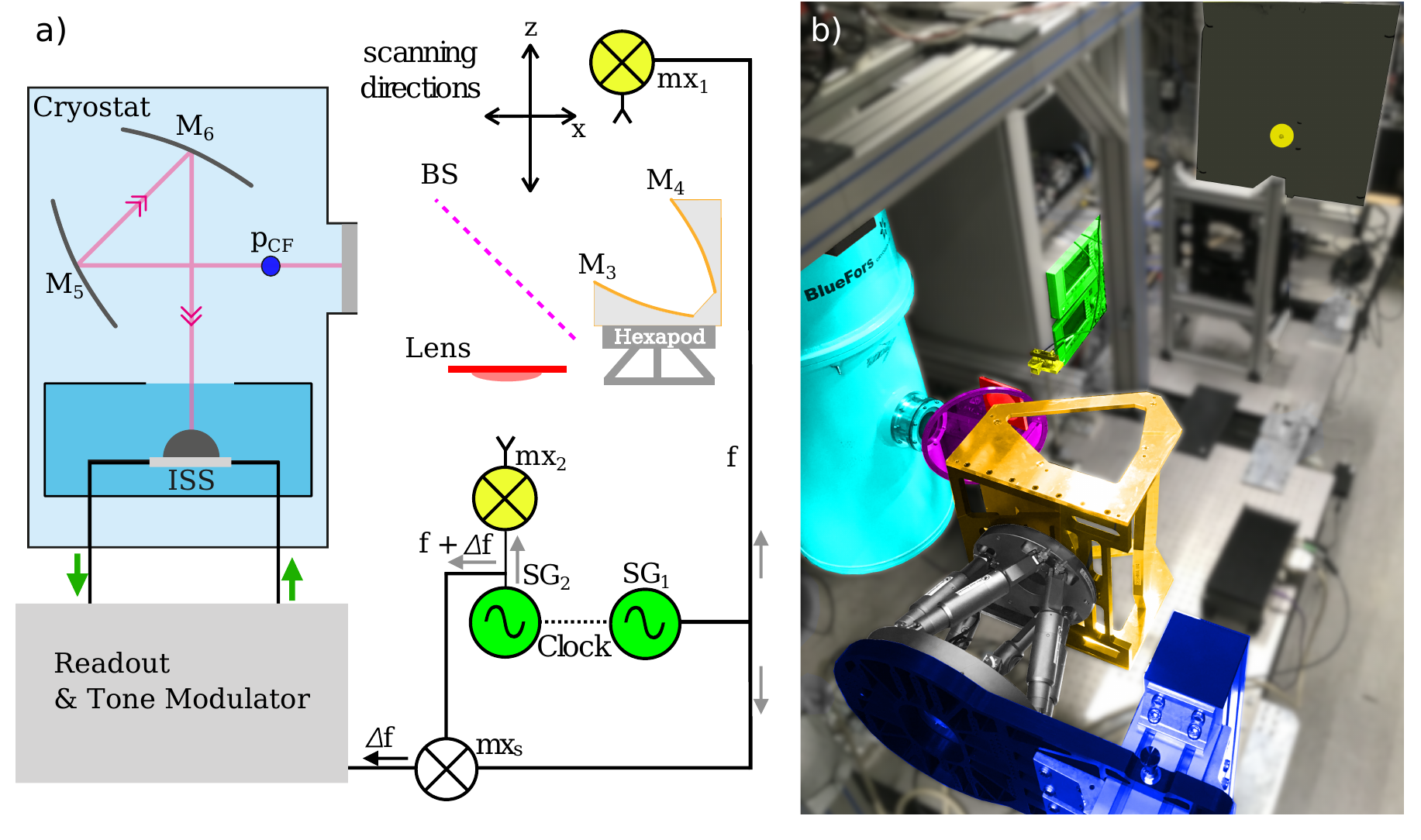}
    \caption{a) Sketch of the HPA measurement setup around the warm focal plane. The sketch is similar to the sketch in Fig.~\ref{fig:setup_HPA}a, the only difference being the addition of the warm optics and hexapod and a change of the scanning directions resulting from the positioning of mx$_1$. b) Photograph of the lab setup, with the scanning plane after the warm optics. The lab setup is essentially the same as Fig.~\ref{fig:setup_HPA}b, except for the addition of the warm optics, which is colored orange. Also, the signal generators can be seen now. Note the $-90^\circ$ rotation around the optical axis coming out of the cryostat window with respect to the optical configuration shown in Fig.~\ref{fig:setup_HPA}b. The lab photograph has the same color coding as the sketch in a). Note that, because the optical axis coming from the warm optics is now oriented along the $y$-axis, the scanning plane is now oriented along the $x$ and $z$-axes. This is also reflected in a).}
    \label{fig:HPA_WO}
\end{figure*}

To measure the misalignment in the warm focal plane, we repeat the HPA measurement around the warm focal plane and calculate the beam offsets in the warm focal plane using the beam offset in the scanning plane and the Gaussian fit. The beam tilts in the scanning plane are identical to those in the warm focal plane and can be used as-is. See Fig.~\ref{fig:HPA_WO}a for a sketch and Fig.~\ref{fig:HPA_WO}b for a photograph of this HPA setup in the laboratory.

\subsection{Optimisation strategy}
\label{subsec:optimisation}
In essence, the optimisation strategy involves simulating the warm optics and hexapod, and performing ray-traces through the warm optics to calculate the optimal hexapod configuration, given some misalignment at the cryostat. All the ray-tracing calculations and simulations of the warm optics are done using the optical simulation software \texttt{PyPO}~\citep{Moerman2023}.

We start by defining a ray representation $B$ of a Gaussian beam~\citep{Crooker2006}, oriented along the $x$-axis, and placing the focus in $\Vec{p}_\mathrm{CF}$. We apply the beam tilt at the cold focal plane to $B$ by orienting $B$ along the measured $\Vec{\hat{\beta}}'_B$.
Then, the rays are propagated through the warm optics and into the warm focal plane where the following cost function is evaluated:

\begin{align}
\label{eq:merit}
    \delta(\Vec{p}_\mathrm{hex}, \Vec{\theta}_\mathrm{hex}) = \frac{|\Vec{p}_B(\Vec{p}_\mathrm{hex}, \Vec{\theta}_\mathrm{hex}) - \Vec{p}_0|}{\Delta{l_\mathrm{min}}} 
    + \frac{\arccos( \Vec{\hat{\beta}}_B(\Vec{p}_\mathrm{hex}, \Vec{\theta}_\mathrm{hex}) \cdot \Vec{\hat{\gamma}}_0)}{\Delta{\alpha_\mathrm{min}}}.
\end{align}

\begin{figure}[ht]
\centering
    \includegraphics[scale=0.7]{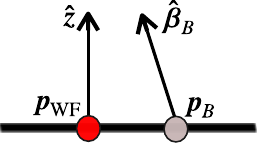}
    \caption{2D sketch of the warm focal plane and the parameters described in Eq.~\ref{eq:merit}. The figure is a zoom of the warm focal plane shown in Fig.~\ref{fig:setup_HPA}, with the misaligned ray direction $\hat{\boldsymbol{\beta}}$ exaggerated. The coordinate system is identical to the one shown in Fig.~\ref{fig:setup_WO}a.}
    \label{fig:WFP_zoom}
\end{figure}

Here, $\Vec{p}_B$ represents the geometric centre of all rays in $B$ evaluated in the warm focal plane, $\Vec{\hat{\beta}}_B$ the mean propagation direction of $B$, $\Vec{p}_\mathrm{hex}$ the simulated hexapod position, and $\Vec{\theta}_\mathrm{hex}$ the simulated hexapod orientation. The beam offset is optimised with respect to $\Vec{p}_0$ and the beam tilt with respect to $\Vec{\hat{\gamma}}_0$. In this case, $\Vec{p}_0=\Vec{p}_\mathrm{WF}$ and $\Vec{\hat{\gamma}}_0=\Vec{\hat{z}}$. The parameters $\Delta{l_\mathrm{min}}$ and $\Delta{\alpha_\mathrm{min}}$ are weights that control the required accuracy in the optimisation. In this work, we adopt the following values: $\Delta{l_\mathrm{min}}=1$ mm and $\Delta{\alpha_\mathrm{min}}=0.1^\circ$. Note that we only consider the beam offset in the warm focal plane and not along the optical axis. This implies that the strategy does not explicitly correct for defocus, which is misalignment along the optical axis, but only for lateral misalignment in the warm focal plane. We find that the inclusion of the focal position in Eq.~\ref{eq:merit}, by means of adding the root-mean-square (RMS) size of the ray-trace beam evaluated in the warm focal plane, divided by a weight of $\Delta\mathrm{RMS}_\mathrm{min} = 1$ mm, does not produce a difference in hexapod configuration or beam RMS size compared to an optimisation without focal position. Including the focus with a $\Delta\mathrm{RMS}_\mathrm{min} = 0.1$ mm results in a reduced beam RMS size, indicating that the focus is now closer to the warm focal plane, but also results in a large beam offset and tilt. This indicates that, at least with some misaligned initial conditions at the cold focus, it is not possible to fully correct for defocusing while at the same time having the beam intersect the warm focus at perpendicular incidence to the warm focal plane. We find that the typical distance along the $z$-axis between the warm focus to the ray-trace beam focus is on the order of $\sim$100 mm, which can easily be corrected for by adjusting $M_2$ of the ASTE telescope along the $z$-axis. Therefore, we do not include the focal position in Eq.~\ref{eq:merit} and optimise by minimising only the beam offset and tilt. For a sketch of the local warm focal plane geometry with optimisation quantities, see Fig.~\ref{fig:WFP_zoom}.

The minimisation of Eq.~\ref{eq:merit} is performed using the differential evolution optimisation algorithm~\citep{Storn_1997}, which is well suited for optimisation problems with several degrees of freedom. The optimal hexapod configuration is then $\Vec{p}_\mathrm{opt}$ and $\Vec{\theta}_\mathrm{opt}$, which are the $\Vec{p}_\mathrm{hex}$ and $\Vec{\theta}_\mathrm{hex}$ corresponding to the lowest value of $\delta$, and can be applied as a correction to the real-world hexapod to align the optical setup. 

As mentioned before, the hexapod and warm optics setup could be misaligned as well, in addition to misalignment at the cryostat window. To find this we first measure the beam center $\Vec{p}^0_B$ in the warm focal plane and propagation direction $\Vec{{\hat{\beta}}}^0_B$ after the warm optics using the HPA method. The superscript `0' indicates that this measurement takes place before the optimisation procedure.
By then optimising Eq.~\ref{eq:merit} with $\Vec{p}_0=\Vec{p}^0_B$ and $\Vec{\hat{\gamma}}_0=\Vec{{\hat{\beta}}}^0_B$, we obtain $\Vec{p}^0_\mathrm{hex}$ and $\Vec{\theta}^0_\mathrm{hex}$, the real-world translational and rotational hexapod offsets contributing to the measured beam misalignment in the warm focal plane.
These hexapod offsets can then be added on top of the $\Vec{p}_\mathrm{hex}$ and $\Vec{\theta}_\mathrm{hex}$ tried by the optimisation in Eq.~\ref{eq:merit}, with $\Vec{p}_0=\Vec{p}_\mathrm{WF}$ and $\Vec{\hat{\gamma}}_0=\Vec{\hat{z}}$, to take into account the real-world hexapod offsets:

\begin{align*}
\Vec{p}_\mathrm{hex} &\xrightarrow{} \Vec{p}_\mathrm{hex} + \Vec{p}^0_\mathrm{hex}, \\
\Vec{\theta}_\mathrm{hex} &\xrightarrow{} \Vec{\theta}_\mathrm{hex} + \Vec{\theta}^0_\mathrm{hex}. \\
\end{align*}

The optimisation then returns the optimised hexapod translation $\Vec{p}_\mathrm{opt}$ and rotation $\Vec{\theta}_\mathrm{opt}$, calculated on top of $\Vec{p}^0_\mathrm{hex}$ and $\Vec{\theta}^0_\mathrm{hex}$.
When applying the optimisation result to the real-world hexapod, care should be taken to only apply $\Vec{p}_\mathrm{opt}$ and $\Vec{\theta}_\mathrm{opt}$, and not $\Vec{p}^0_\mathrm{hex}$ and $\Vec{\theta}^0_\mathrm{hex}$.

Then, an HPA measurement around the warm focal plane can be performed to assess the new beam center $\Vec{p}^1_B$ and propagation direction $\Vec{\hat{\beta}}^1_B$. Here, the `1' superscript indicates that these misalignments are measured after the optimised configuration is applied to the real-world hexapod.
If significant residual misalignment is present, an iterative approach is taken. For this, we first generate a characterisation of the warm optics (see Appendix~\ref{app:DoFs}) to see how the hexapod degrees-of-freedom (DoFs) affect the beam misalignment. Using this characterisation, we can select hexapod DoFs to include in the optimisation and hexapod offset finding.

The optimisation strategy can directly be applied at the telescope. This involves recreating the experimental setup shown in Fig.~\ref{fig:HPA_WO}b inside the lower cabin. The mixer mx$_1$ needs to be placed in the upper cabin (see Fig.~\ref{fig:setup_WO}a) and placed such that it faces $M_3$. This adaptability makes the strategy versatile, as it can be applied in a variety of contexts.

\subsection{Experimental setup}
\setstcolor{red}
\label{subsec:expsetup}
To test the alignment procedure, we recreate the lower cabin setup at ASTE (see Fig.~\ref{fig:HPA_WO}b) in the laboratory. Because the laboratory ceiling is not high enough to place mx$_1$ above the warm optics, the warm optics and hexapod are rotated by $-90^\circ$ around the $x$-axis. This places the warm focal plane parallel to the $xz$ plane and places the optical axis coming out of the warm optics parallel the $y$-axis. Consequently, the beam tilts in the warm focal plane are now defined and measured with respect to the $x$ and $z$-axes, instead of the $x$ and $y$-axes. Also, the 30$^\circ$ tilt of the cryostat around the $x$-axis present in the ASTE lower cabin setup (see Fig.~\ref{fig:setup_WO}b) is not applied to the laboratory cryostat.

For the HPA measurements in front of the cryostat window, we place the scanning plane about 30 cm after the cold focal plane along the positive $x$-axis. For the HPA measurements after the warm optics, we place the scanning plane about 25 cm after the warm focal plane along the $y$-axis. 

We manually measure the translation of the scanning plane from the position in front of the cryostat window to the position after the warm optics. Because we know the distance between $\Vec{p}_\mathrm{CF}$ and the scanning plane center after the cryostat window, the scanning plane translation can be used to obtain the absolute position of the scanning plane center after the warm optics with respect to $\Vec{p}_\mathrm{CF}$, and hence $\Vec{p}_\mathrm{WF}$. This allows us to place the measured beam center after the warm optics in the co-ordinate system used for the optimisation described in Sect.~\ref{subsec:optimisation} and to quantitatively assess the offset between $\Vec{p}_\mathrm{WF}$ and the measured beam focus in the warm focal plane during the iterative part of the optimisation.

In order to simulate propagation of the measured beam patterns through the Cassegrain setup at ASTE, we rotate the beam patterns back by $90^\circ$ around the $x$-axis, so that the scanning plane normal is oriented along the $z$-axis.
Then, we take the fitted beam focus position, averaged across the HPA frequency range, and assume that this position overlaps with the Cassegrain focus of ASTE. This allows us to place the scanning plane in a model of ASTE and simulate the propagation of the measured beam patterns using \texttt{PyPO}, while keeping the distance between $M_2$ and the warm focal plane fixed at the design distance. At the real telescope, this corresponds to focus correction by $M_2$, as described in Sect.~\ref{subsec:optimisation}, such that the Cassegrain focus overlaps with the fitted beam focus.

\section{Results}
\label{sect:results}

\subsection{Reduction of beam misalignment}
\label{subsec:reduction}
The first result we find is a reduction of the beam misalignment. We measure the beam tilt coming out of the cryostat to be $\theta'_y\approx-0.4^\circ$ and $\theta'_z\approx-3.1^\circ$, around the $y$ and $z$-axes, respectively. The misalignment around the $z$-axis is consistent with a rotation in the cryostat mounting, which we measured to be $\sim -2.5^\circ$.
We apply the iterative optimisation procedure as described. Because the hexapod DoF characterisation (see Appendix~\ref{app:DoFs}) indicated that rotational and translational DoFs are degenerate (that is, they have the same effect on the beam direction after the warm optics), we only included translational DoFs in the iterative optimisation.

\begin{figure}
\centering
    \includegraphics[width=0.5\textwidth]{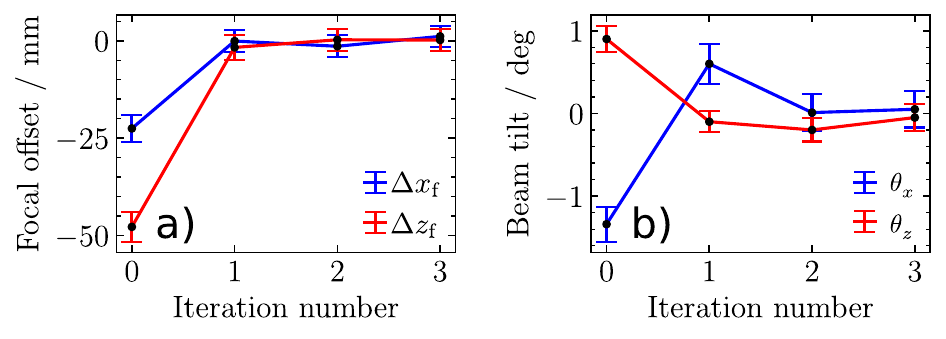}
    \caption{a) Decrease in focal offset, defined as $\Vec{p}_B-\Vec{p}_\mathrm{WF}$. The red data points correspond to offsets along the $z$-axis and the blue data points to offsets along the $x$-axis. b) Decrease in beam tilts around the $x$ and $z$-axes. Both the focal offset and beam tilt are illustrated as function of iteration number. Here, `0' corresponds to the setup in the hexapod home position and `1' to the initial optimised hexapod configuration. Subsequent iteration numbers corresponds to steps of the iterative procedure proper. The lines are drawn between the data points to highlight the tendencies of the iterations.}
    \label{fig:opt_res}
\end{figure}

The alignment procedure required two iterations on top of the initial optimisation in total and took roughly three hours. This corresponds to five HPA measurements in total, one in front of the cryostat and four around the warm focal plane. Most of the total time was spent on the HPA measurements, which was about 30 minutes per measurement of which 10 minutes was spent uploading and downloading measurement data. The numerical optimisation and hexapod offset finding were negligible in the time budget, each taking about 10 seconds to complete. This is due to the efficient multi-threaded differential evolution implementation in SciPy~\citep{Virtanen2020} and the restriction of the DoF space of the hexapod to translational DoFs.
The results for the warm focal plane beam offset and beam tilt can be found in Fig.~\ref{fig:opt_res}.

It can be seen from Fig.~\ref{fig:opt_res} that the beam offset in the warm focal plane is already reduced significantly after the initial optimisation. The final beam offset is on the order of 1 mm, which is an order of magnitude smaller than the initial beam offset. The beam tilt converges slower, but reaches an acceptable tilt smaller than 0.1$^\circ$ after the second iteration.

\subsection{Increase in telescope efficiency}
\label{subsec:efficiency}
\begin{figure*}
\centering
    \includegraphics[scale=0.5]{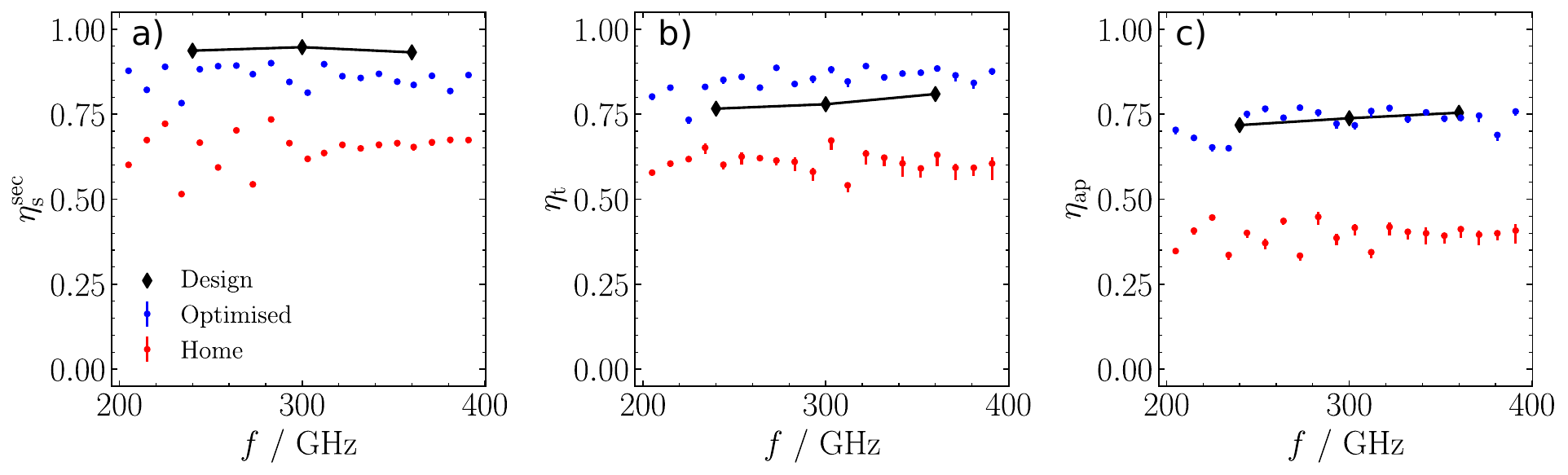}
    \caption{Calculated telescope efficiencies for the ISS at ASTE. a) Spillover efficiency $\etass$ calculated on $M_2$. b) Taper efficiency $\etat$ calculated in the aperture of $M_1$. c) Aperture efficiency $\etaap$, defined as $\etass \etat$. Blue dots represent the optimised configuration, red dots the home configuration and black diamonds the design values of an ideal setup. The errorbars are calculated from the error in the Gaussian fit used to determine the beam misalignment.}
    \label{fig:effs}
\end{figure*}

We find that the telescope efficiency of the instrument is substantially increased by the optimisation. The measured PA beam patterns are propagated through a model of ASTE using physical optics~\citep{Balanis1999-qe} and the efficiency terms in Appendix~\ref{app:effs} are calculated. The results are shown in Fig.~\ref{fig:effs}.

It can be seen that the optimised setup performs considerably better than the misaligned setup in the hexapod home configuration, for all three efficiencies, across all measured frequencies. The secondary spillover efficiency $\etass$ is slightly lower than the design value, whilst the taper efficiency $\etat$ is slightly higher. The decrease in $\etass$ could be due to a higher illumination level on the edge of $M_2$, which also explains the increase in $\etat$. Nevertheless, the aperture efficiency $\etaap$ is roughly consistent with the design value and considerably better than the misaligned values. It should be noted that, for both the optimised and misaligned setup, Ruze losses~\citep{Ruze1966} due to surface roughness of the ASTE Cassegrain are not taken into account. 

We calculate the mean $\etaap$ across the frequency range for the optimised configuration to be $\etaap=0.73 \pm 0.04$. In comparison, the home (misaligned) configuration has a mean $\etaap=0.39 \pm 0.03$. To put this into perspective, we can, for the optimised and home configuration, calculate the ratio of necessary observation times to obtain the same signal-to-noise ratio (S/N):

\begin{equation}
\label{eq:Rt}
    R_t = \frac{t'}{t} \propto \left( \frac{\etaap}{\etaap'} \right)^2,
\end{equation}

where the primed quantities denote either the optimised or home configuration. Note that Eq.~\ref{eq:Rt} is valid for point sources only. We find that our optimised hexapod configuration results in a decrease in required observation time of a factor $R_t = 3.5 \pm 0.7$. This can have a significant impact, especially for long observations that require high S/N for faint targets.

\subsection{Far-field analysis}
\label{subsec:ff}
\begin{figure}
\centering
    \includegraphics[width=0.5\textwidth]{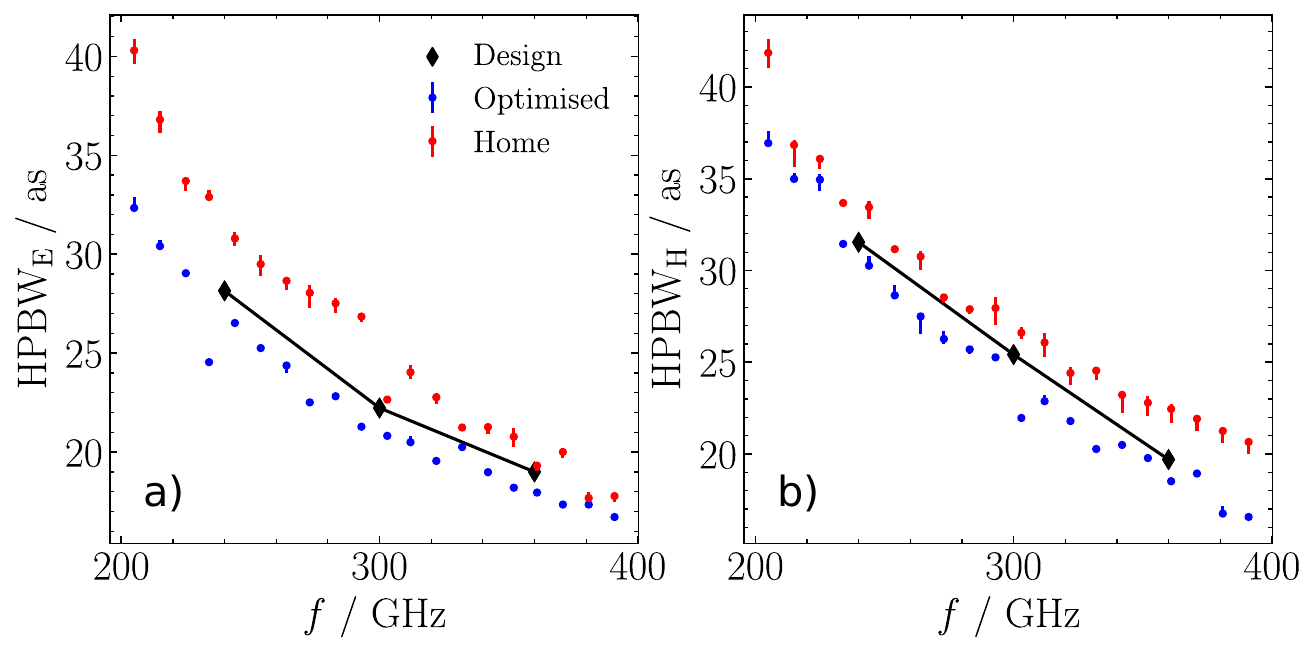}
    \caption{Calculated HPBWs for the home and optimised configuration as function of signal frequency. a) HPBWs in the E-plane. b) HPBWs in the H-plane.}
    \label{fig:eff_mb_hpbw}
\end{figure}

We propagate the measured beam patterns to the telescope far-field using \texttt{PyPO} and analyse the results. In Fig.~\ref{fig:eff_mb_hpbw} we show the effect of the optimisation on the half-power beamwidths (HPBWs) in the E and H-plane. Here, we define the E-plane to lie along the semi-minor axis of the far-field main beam and the H-plane to lie along the semi-major axis. We find that the HPBWs of the optimised setup are decreased with respect to the misaligned setup and are slightly smaller than the design values across all measured frequencies. This is consistent with a higher edge illumination on the secondary, which was already hypothesised as an explanation for the lower $\etass$ and larger $\etat$. By design, the primary reflector of the telescope is under-illuminated at the edges and by slightly increasing the illumination level towards the outer sections on the primary, we slightly decrease our beam size.

\begin{figure}
\centering
    \includegraphics[width=0.5\textwidth]{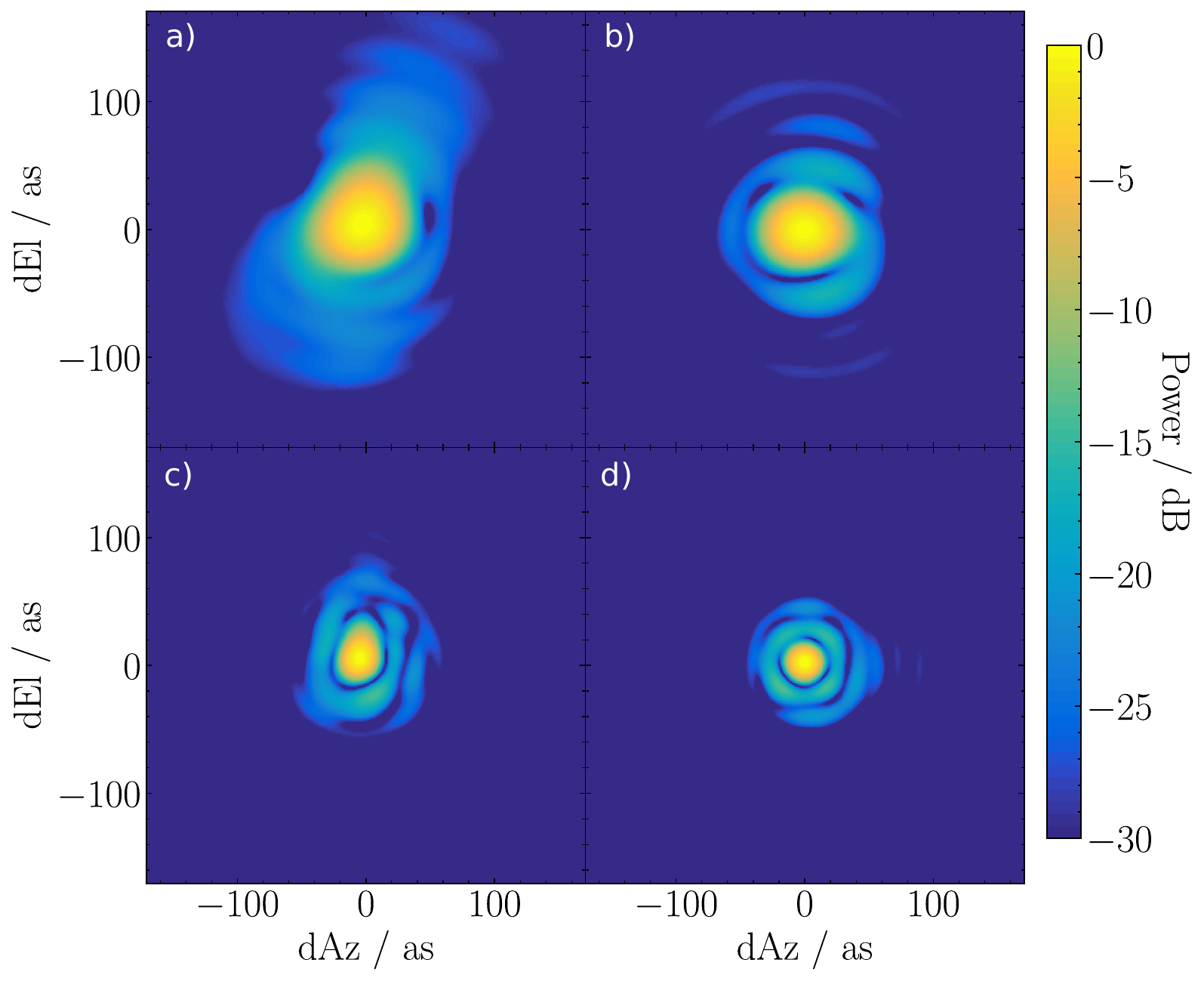}
    \caption{2D far-field beam patterns on-sky. a) 205 GHz, hexapod in home configuration. b) 205 GHz, hexapod in optimised configuration. c) 391 GHz, hexapod in home configuration. d) 391 GHz, hexapod in optimised configuration. All beam patterns are centered such that the beam centers are at $\mathrm{dAz}=\mathrm{dEl}=0$ as.}
    \label{fig:ff}
\end{figure}

Lastly, we perform a qualitative comparison of the far-field beam patterns. We restrict the discussion to the lowest and highest measured frequencies, at 205 and 391 GHz respectively. In Fig.~\ref{fig:ff} we compare the 2D far-field beam patterns.

It is clear that the optimisation reduces the beam ellipticity. This is consistent with Eq.~\ref{eq:merit}, because the optimisation criteria favour a centrally illuminated secondary. With the hexapod in home configuration, the beam pattern is asymmetrically illuminating the secondary reflector and, therefore, primary aperture, giving rise to elliptical beam patterns in the far-field. This also indicates that the home configuration suffers from lower $\etass$, as a smaller fraction of the beam pattern is intercepted by the secondary. This finding supports the observation from Fig.~\ref{fig:effs} and is another indication of the veracity of the alignment procedure. 

\section{Conclusions}
We have developed an alignment procedure for the DESHIMA 2.0 wideband sub-mm ISS, utilising an optimisation strategy together with a hexapod and a modified Dragonian dual reflector. We used the novel HPA measurement technique to obtain the phase and amplitude beam patterns of the instrument and used these to quantitatively measure the misalignment in our optical chain. Because we directly utilise the sub-mm beam of the ISS instead of a more conventional laser guide, we can accurately quantify the misalignment in the optical system. Then, using the optimisation strategy, we obtained a hexapod configuration which successfully mitigated the misalignment that was present. This was verified by laboratory measurements that show that the beam misalignment with the optimised hexapod configuration was sufficiently reduced. The calculated aperture efficiencies of the aligned ISS at the ASTE telescope are improved with respect to the misaligned case. Moreover, the calculated far-field after the telescope shows improvement in the form of reduced main beam size and more symmetric beam shapes. All these findings are mutually consistent and support the veracity of the alignment procedure.

The proposed alignment procedure can be directly employed at the telescope by putting the HPA setup in the lower cabin, emulating Fig.~\ref{fig:HPA_WO}b. The scanning plane itself can be placed in the ASTE upper cabin. Although this would be a complex effort due to the limited space present in the telescope lower and upper cabin, it is not impossible. For example, the scanning plane movement stage could be attached to the top of the hexaport support structure (see Fig.~\ref{fig:setup_WO}b) with the stage protruding upward into the upper cabin, allowing the scanning source itself to be located in the upper cabin and eliminating the need for additional support structures in the upper cabin.
The importance of the alignment procedure can be appreciated in the context of the upcoming science verification campaign of DESHIMA 2.0. One science case of the campaign, rapid redshift surveys of dusty star-forming galaxies~\citep{Rybak_2022}, predicts a 400 hour observation period to yield robust redshifts of the targets. This assumes a well-aligned system and if misalignment is present, the required observation time will increase significantly. 

The alignment procedure is also of interest for future extensions of the single-pixel ISS towards an ISS-based IFU. The cost function in Eq.~\ref{eq:merit} is designed to align the ISS to the optical axis of the telescope. Because the HPA technique can be used to measure the beam patterns of the pixels in the IFU individually, alignment information can be extracted for each pixel individually. Then, the optimisation procedure in this work can be used again by using a cost function more suitable for multi-pixel instruments.

\begin{acknowledgements}
This work was supported by the European Union (ERC Consolidator Grant No. 101043486 TIFUUN). Views and opinions expressed are however those of the authors only and do not necessarily reflect those of the European Union or the European Research Council Executive Agency. Neither the European Union nor the granting authority can be held responsible for them. TT was supported by the
MEXT Leading Initiative for Excellent Young Researchers (Grant No.
JPMXS0320200188).

\end{acknowledgements}

%
%
\bibliographystyle{aa} 
\bibliography{refs}

\begin{appendix}
\section{Characterising hexapod DoFs}
\label{app:DoFs}
In order to link hexapod DoFs to beam misalignment parameters, a characterisation can be performed. This involves performing ray-traces through the warm optics, from $\Vec{p}_\mathrm{CF}$ into the warm focal plane. We do not apply $\Vec{\hat{\beta}}'_B$ to the ray-trace beam for the characterisation. For each ray-trace, we adjust the hexapod configuration along a single degree of freedom (DoF), whilst keeping the other DoFs fixed at home position, i.e. at zero translation/rotation. In this way, we can characterise the effect a single DoF has on the simulated beam offset and tilt in the warm focal plane. Then, specific DoFs that couple to the measured beam misalignment can be selected in the matching or optimisation run.
Here we present an example in the context of the optical system treated in this work. We show a characterisation for $Z$, the hexapod DoF which lies along the $z$-axis in Fig.~\ref{fig:setup_WO}a. 

\begin{figure}
\centering
    \includegraphics[width=0.45\textwidth]{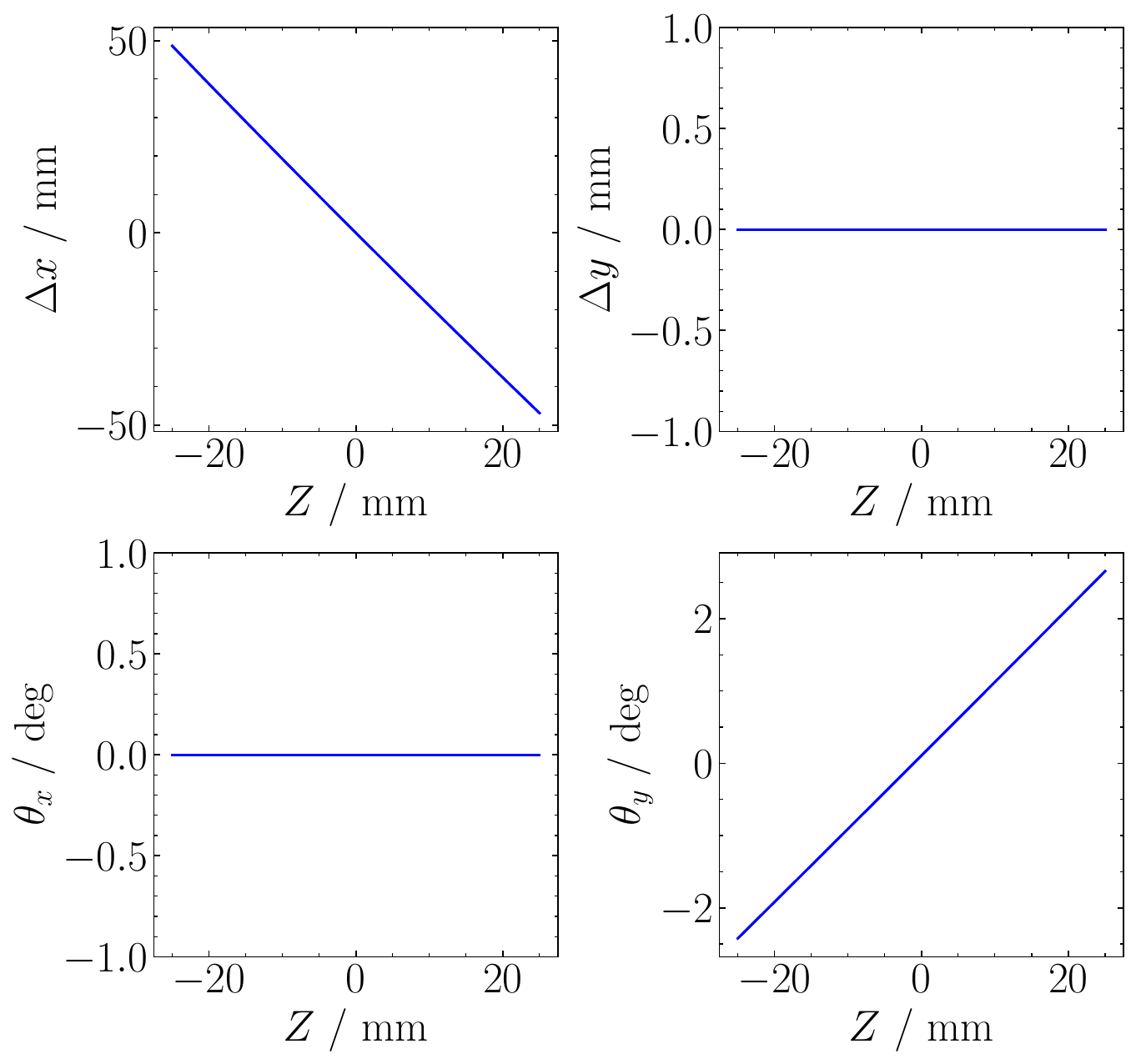}
    \caption{Beam misalignment parameters versus the hexapod DoF along the $z$-axis, in millimeters. The ray-trace beam is evaluated in the warm focal plane, according to Fig.~\ref{fig:setup_WO}a. a) Beam offset $\Delta x$ along the $x$-axis in mm. b) Beam offset $\Delta y$ along the $y$-axis in mm. Both offsets are calculated with respect to $\Vec{p}_\mathrm{WF}$. c) Beam tilt $\theta_x$ around the $x$-axis in degrees. d) Beam tilt $\theta_y$ around the $y$-axis in degrees.}
    \label{fig:z_char}
\end{figure}

It is evident from Fig.~\ref{fig:z_char} that this particular DoF strongly couples to the beam offset along the $x$-axis and the beam tilt around the $y$-axis. If a measured beam is misaligned in these two parameters, this DoF would be included in the matching-optimising steps of the iterative optimisation procedure. 
It does not couple to the beam offset along the $y$-axis and the beam tilt around the $x$-axis. This indicates that, if misalignment is found in any of the non-coupling parameters, another DoF needs to be considered for inclusion in the matching-optimising.

\section{Telescope efficiencies}
\label{app:effs}
Here we give the used expressions for $\etass$, $\etat$ and $\etaap$. The secondary spillover $\etass$ is calculated over the secondary aperture and is given by:
\begin{equation}
    \label{eq:spill}
    \etass = \frac{| \int_{S_\mathrm{a}} E^\star_i E_i \mathrm{d}{S_\mathrm{a}} |^2}{\int_{S_\mathrm{a}} |E_i|^2 \mathrm{d}{S_\mathrm{a}} \int_{S'_\mathrm{a}} |E_i|^2 \mathrm{d}{S'_\mathrm{a}}},
\end{equation}
where $E_i$ denotes the electric field illuminating the extended reflector aperture plane $S'_a$, which is defined as the spatially extended version of the secondary reflector aperture $S_a$. In practice, we oversize $S'_a$ to have a radius three times that of $S_a$, so that we capture sufficient spillover radiation. The $\star$ denotes complex conjugation. This efficiency is a measure of how much illumination is intercepted by the secondary reflector. Because the spillover losses on the primary reflector are negligible in both the misaligned and optimised configuration, these are not discussed.

The second efficiency we discuss is the taper efficiency $\etat$. This efficiency is a measure of the uniformity of the amplitude and phase patterns of the beam in the primary aperture plane $P_\mathrm{a}$ and is calculated as follows:
\begin{equation}
\label{eq:taper}
    \etat = \frac{1}{\Sigma}\frac{| \int_{P_\mathrm{a}} E_P \mathrm{d}{P_\mathrm{a}} |^2}{\int_{P_\mathrm{a}} |E_P|^2 \mathrm{d}{P_\mathrm{a}} },
\end{equation}
where $\Sigma$ denotes the physical surface area of ${P_\mathrm{a}}$, $E_i$ the electric field illuminating ${P_\mathrm{a}}$. We always orient the aperture in which we evaluate $\etat$ such that the aperture normal is parallel to the telescope pointing. In this way, we calculate $\etat$ in the direction of maximum directivity.
With these two efficiencies, we can estimate the aperture efficiency $\etaap$~\citep{goldsmith1998}:
\begin{equation}
\label{eq:aper}
    \etaap \approx \etass \etat.
\end{equation}

\section{Directivities of aligned ISS}
\begin{figure}
\centering
    \includegraphics[width=0.3\textwidth]{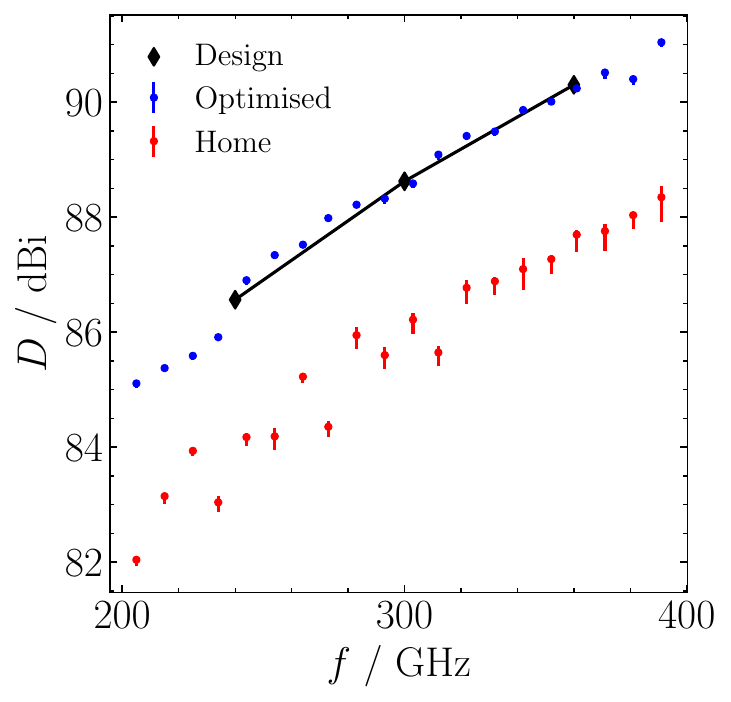}
    \caption{Directivities of the ISS after the ASTE telescope. In red, directivities with the hexapod in home position. In blue, directivities with the optimised hexapod position. In black, design directivities for an aligned ISS.}
    \label{fig:directivities}
\end{figure}

The increase in performance after aligning is also reflected in the increase of directivity of the instrument beam. The directivity depends on $\etaap$ and is given by:
\begin{equation}
    D_\mathrm{dBi} = \log_{10}\left( \frac{4 \pi}{\lambda^2} \etaap \Sigma \right).
\end{equation}
Here, $D_\mathrm{dBi}$ is the directivity in decibels (with respect to the isotropic radiator), $\lambda$ is the wavelength of radiation and $\Sigma$ is the surface area of the primary aperture.
Because of the prevalence of this metric in certain fields, we include the directivities in Fig.~\ref{fig:directivities} for completeness.

\end{appendix}
\end{document}